\documentclass[prl,aps,floats,amsmath,amssymb,twocolumn,superscriptaddress,preprintnumbers,nofootinbib]{revtex4}
\usepackage{graphicx}
\usepackage{dcolumn}
\usepackage{bm}
\usepackage{hyperref}
\usepackage{epstopdf}
\usepackage{color}

\def\lsim{\raisebox{-4pt}{$\,\stackrel{\textstyle{<}}{\sim}\,$}}
\def\gsim{\raisebox{-4pt}{$\,\stackrel{\textstyle{>}}{\sim}\,$}}

\def\beq{\begin{equation}} 
\def\eeq{\end{equation}} 
\def\bea{\begin{eqnarray}} 
\def\eea{\end{eqnarray}} 
\def\ben{\begin{enumerate}} 
\def\een{\end{enumerate}} 
\def\ie{{\it i.e.}}

\def\lsim{\mathrel{\raise.3ex\hbox{$<$\kern-.75em\lower1ex\hbox{$\sim$}}}} 
\def\gsim{\mathrel{\raise.3ex\hbox{$>$\kern-.75em\lower1ex\hbox{$\sim$}}}} 
\def\ifmath#1{\relax\ifmmode #1\else $#1$\fi}

\def\simlt{\stackrel{<}{{}_\sim}}
\def\simgt{\stackrel{>}{{}_\sim}}

\begin{document}
\DeclareGraphicsExtensions{.jpg,.pdf,.mps,.png,}

\preprint{ANL-HEP-PR-10-48}
\preprint{EFI-10-22}
\title{Dark Light Higgs}

 \author{Patrick Draper}
\affiliation{Enrico Fermi Institute, 
University of Chicago, Chicago, IL
60637, USA}

\author{Tao Liu}
\affiliation{Enrico Fermi Institute, 
University of Chicago, Chicago, IL
60637, USA}
\affiliation{Department of Physics, University of California,
Santa Barbara, CA 93106, USA}

\author{Carlos E.M.~Wagner}
\affiliation{Enrico Fermi Institute, 
University of Chicago, Chicago, IL
60637, USA}
\affiliation{HEP Division, Argonne National Laboratory, 9700 Cass Ave., Argonne, IL 60439, USA}
\affiliation{KICP and Dept. of Physics, Univ. of Chicago, 5640 S. Ellis Ave., Chicago IL 60637, USA}

\author{Lian-Tao~Wang}
\affiliation{Department of Physics, Princeton University, Princeton, NJ 08540, USA}

\author{Hao Zhang}
\affiliation{Enrico Fermi Institute, 
University of Chicago, Chicago, IL
60637, USA}
\affiliation{Department of Physics and State Key Laboratory of Nuclear Physics
and Technology, Peking University, Beijing 100871, China}

\begin{abstract}

We study a limit of the nearly-Peccei-Quinn-symmetric Next-to-Minimal Supersymmetric Standard Model possessing novel Higgs and dark matter (DM) properties. In this scenario, there naturally co-exist three light singlet-like particles: a scalar, a pseudoscalar, and a singlino-like DM candidate, all with masses of order 0.1-10 GeV. The decay of a Standard Model-like Higgs boson to pairs of the light scalars or pseudoscalars is generically suppressed, avoiding constraints from collider searches for these channels. For a certain parameter window annihilation into the light pseudoscalar and exchange of the light scalar with nucleons allow the singlino to achieve the correct relic density and a large direct detection cross section consistent with the CoGeNT and DAMA/LIBRA preferred region simultaneously. This parameter space is consistent with experimental constraints from LEP, the Tevatron, $\Upsilon$- and flavor physics.

\end{abstract}

\maketitle

The Next-to-Minimal Supersymmetric Standard Model (NMSSM) is a well-motivated extension of the Minimal Supersymmetric Standard Model (MSSM) by a gauge-singlet chiral superfield $\mathbf{N}$, designed to solve the $\mu$-problem of the MSSM. 
Its superpotential and soft supersymmetry-breaking terms in the Higgs sector are
\begin{align}
\mathbf{W}&=\lambda \mathbf{N}\mathbf{H_u}\mathbf{H_d} + \frac{1}{3}\kappa \mathbf{N}^3, \nonumber \\
 V_{\it soft} &= {m^2_{H_d}} |H_d|^2 + {m^2_{H_u}} |H_u|^2
+{m^2_N}|N|^2 \nonumber \\
 & - (\lambda A_{\lambda} H_u H_d N + h.c.)+\left(\frac{\kappa}{3} A_{\kappa} N^3 +
h.c.\right). 
\label{higgspot}
\end{align}
Here $H_d$, $H_u$ and $N$ denote the neutral Higgs bosons
corresponding to $\mathbf{H_d}$, $\mathbf{H_u}$ and $\mathbf{N}$,
respectively.

In this work, we examine an NMSSM limit 
given by two conditions. The first one is $\kappa\ll\lambda$ which is protected by an approximate Peccei-Quinn (PQ) symmetry. It is well-known that a light pseudoscalar $a_1$ will be generated by the spontaneous breaking of such a $U(1)$ symmetry (the phenomenology on a light $a_1$ has been thoroughly studied in the R-symmetry limit~\cite{Dobrescu:2000jt,Dermisek:2005ar}). As noted in~\cite{Ciafaloni:1997gy}, at tree level the PQ limit implies an upper bound on the lightest scalar mass $m_{h_1}$ approximately proportional to $\lambda^2$. Here we address the further limit of $\lambda\simlt 0.1$, 
leading to the simultaneous emergence of a light singlet-like scalar $h_1$ and a light singlino-like lightest superpartner $\chi_1$. For mildly small values of $\lambda~(\lambda > 0.05)$ studied in this letter, typically $\lambda (\Lambda_{\rm GUT}) \sim {\mathcal O}(0.1)$, a natural order for a perturbative parameter. We stress that this scenario differs from the light $a_1$ case of~\cite{Dobrescu:2000jt,Dermisek:2005ar}, in that $h_1$, $a_1$, and $\chi_1$ are all of order $0.1- 10$ GeV. It also differs in that   
decays of the Standard Model (SM)-like Higgs boson to $h_1h_1$ and $a_1a_1$ pairs are generically suppressed. Thus $h_1$ and $a_1$ are hidden from four-fermion searches at LEP~\cite{Schael:2006cr} and the Tevatron~\cite{Abazov:2009yi} designed to test a light $a_1$ scenario. 
Meanwhile, due to annihilation into $a_1$ and exchange of $h_1$, for a certain window of the parameters, the correct relic density and a large spin-independent (SI) direct detection cross section consistent with the CoGeNT and DAMA/LIBRA preferred region can be achieved for the DM candidate $\chi_1$. Therefore, we refer to this limit as the ``Dark Light Higgs'' (DLH) scenario.

\begin{figure}[ht]
\setlength{\belowcaptionskip}{-20pt}
\begin{center}
\includegraphics[width=0.45\textwidth]{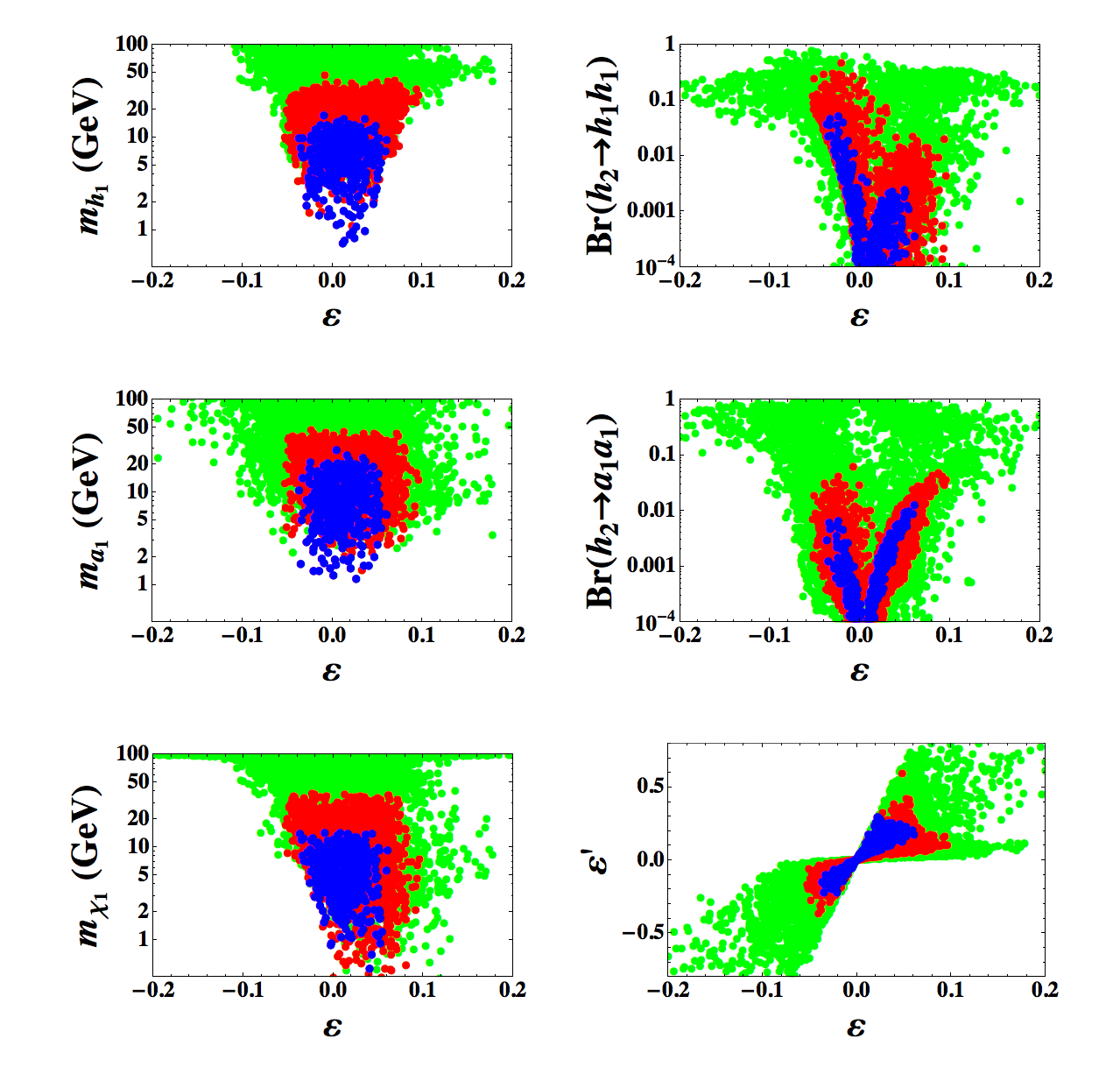}
\caption{Masses of $h_1$ (\textit{top-left}), $a_1$ (\textit{middle-left}), and $\chi_1$ (\textit{bottom-left}); branching ratios of $h_2$ into $h_1h_1$ (\textit{top-right}) and $a_1a_1$ (\textit{middle-right}), and correlation between $\varepsilon$ and $\varepsilon'$ (\textit{bottom-right}). Points are taken randomly from the ranges $5\leq\tan\beta\leq 50$, $0.05\leq\lambda\leq 0.5$, $0.0005\leq\kappa\leq0.05$, $-0.8\leq\varepsilon' \leq 0.8$, $-40\mbox{ GeV}\leq A_\kappa\leq 0$, and $0.1\mbox{ TeV}\leq\mu\leq1\mbox{ TeV}$. (As an illustration, we assume soft squark masses of 1 TeV, slepton masses of 200 GeV, $A_{u,d,e}$  parameters of 750 GeV, and bino, wino and gluino masses of 100, 200 and 660 GeV, respectively, for all numerical analyses in this letter.) Green points cover the whole scan range, red points correspond to $\lambda<0.30$, $\kappa/\lambda<0.05$ and $\mu<400$ GeV, and blue points 
correspond to $\lambda<0.15$, $\kappa/\lambda<0.03$ and $\mu<250$ GeV.  }
\label{massfig}
\end{center}
\end{figure}

We begin with an analysis of the light spectrum in the DLH scenario.  For convenience we define two parameters  
\begin{align}
\varepsilon\equiv \frac{\lambda \mu}{m_Z}   \varepsilon',  \ \  \varepsilon' \equiv \frac{A_\lambda}{\mu\tan\beta} -1
\label{ourlimit}
\end{align}
with $\mu\equiv\lambda\langle N\rangle$. 
$\varepsilon$ has an impact on Higgs physics, as exhibited in FIG.~\ref{massfig}.
In the first column of FIG.~\ref{massfig} we plot $m_{h_1,a_1,\chi_1}$ against $\varepsilon$ for a random scan as defined in the figure caption. NMSSMTools 2.3.1 and MicrOMEGAS 2.4.Q~\cite{NMSSMTools,Belanger:2010gh} are our analysis tools used in this letter.

The scan results in FIG.~\ref{massfig} can be understood analytically as follows. Because of the spontaneous breaking of the approximate PQ symmetry, $a_1$ is a pseudo-Goldstone boson and its small mass  
$m^2_{a_1}\approx-3\kappa A_{\kappa}\mu/\lambda$ 
is protected. 
For $\chi_1$, $\kappa\ll\lambda\ll1$ implies that it is dominantly singlino and its mass is 
$m_{\chi_1}\approx v^2\lambda^2\sin2\beta /\mu+ 2\kappa\mu/\lambda$,
where $v=174$ GeV and $\tan\beta\equiv\langle H_u\rangle/\langle H_d\rangle$. For $\lambda\lesssim 0.1$, $\mu$ of order a few hundred GeV, and $\kappa/\lambda$ on the order of a few percent, $m_{\chi_1}$ drops below 10 GeV.

More interesting is the CP-even spectrum. For analytic convenience we consider moderate $\tan\beta$, although the qualitative properties of the figures are also present for lower $\tan\beta$. 
In the small $\lambda$ + PQ limit $h_1$ has a mass  
\begin{align}
(m_{h_1}^2)_{\rm tree}\approx -4v^2\varepsilon^2 +\frac{4v^2\lambda^2}{\tan^2\beta} + \frac{\kappa A_\kappa \mu}{\lambda}+\frac{4\kappa^2\mu^2}{\lambda^2}.
\label{epsilon}
\end{align}
at tree level. The heaviest state is strongly down-type, with a mass $m_{h_3}^2\simeq m_{H_d}^2\simeq A_{\lambda}^2$ (where the minimization condition for $v_d$ is used), and the middle state is SM-like.

The $h_1$ mass is also lifted by quantum corrections, and
the strong singlet-like nature of $h_1$ suppresses contributions from all particles running in the loop except Higgs bosons and Higgsinos.  
Setting $\varepsilon\rightarrow 0$ for these loop diagrams,  
we find an uplifted singlet mass in the $\overline{MS}$ scheme 
\begin{align}
\Delta m_{h_1}^2\approx \frac{\lambda^2\mu^2}{2\pi^2} \log{\frac{\mu^2\tan\beta^3}{m_Z^2}}.
\label{msloop}
\end{align}
Fixing all other parameters, the upper bound on $m_{h_1}^2$ is achieved for $\varepsilon\rightarrow 0$ and is lowered to about or below $10$ GeV in the small $\lambda$ + PQ limit.

On the other hand, increasing $\varepsilon$ rapidly decreases $m_{h_1}$. Vacuum stability (\ie, $(m_{h_1}^2)_{\rm tree} + \Delta m_{h_1}^2 \ge 0$) indicates an upper bound on $\varepsilon^2$ 
\begin{align}
\varepsilon_{max}^2&\approx\frac{1}{4v^2 }\left (\frac{4\lambda^2v^2}{\tan^2\beta} + \frac{\kappa A_\kappa\mu}{\lambda}   +\frac{4\kappa^2\mu^2}{\lambda^2}+ \Delta m_{h_1}^2\right).
\label{eps0}
\end{align}
In the small $\lambda$ + PQ limit and for natural values of $\mu$, $|\varepsilon_{max}|$ is small.  This fact will be relevant for collider constraints discussed below.
The right-bottom panel of FIG.~\ref{massfig} also shows that $A_\lambda$ is usually close to $\mu \tan\beta$ for blue points, so we will take a smaller range of $\epsilon'$ in our DM analysis.

The tree-level mixing parameters of the light scalar are
\begin{align}
S_{1d}\approx\frac{v}{\mu\tan\beta}\bigg(\lambda +\frac{2\varepsilon\mu}{m_Z}\bigg), \ \  
S_{1u}\approx\frac{2 v\varepsilon}{m_Z},
\label{singmix}
\end{align}
indicating a mostly singlet/down admixture in the limit $\varepsilon\rightarrow 0$ and an approximately pure singlet (\ie, $S_{1s} \to 1$) in the further limit of small $\lambda$ or large $\tan\beta$.

There are three main processes by which present experiments potentially constrain this scenario: (1) decays of the SM-like Higgs $h_2$ through $h_1h_1$ and $a_1a_1$, (2) $\Upsilon (ns)$  decays to $\gamma h_1$ or $\gamma a_1$, and (3) flavor physics.

\begin{figure}[h]
\setlength{\belowcaptionskip}{-8pt}
\includegraphics[width=0.34\textwidth]{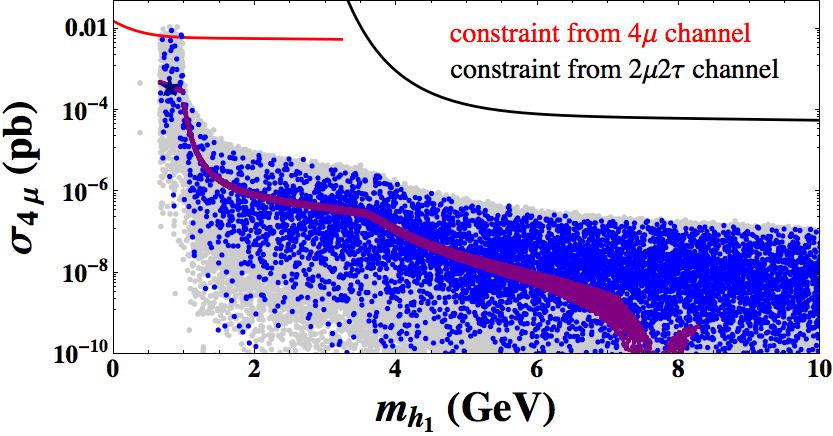}
\includegraphics[width=0.34\textwidth]{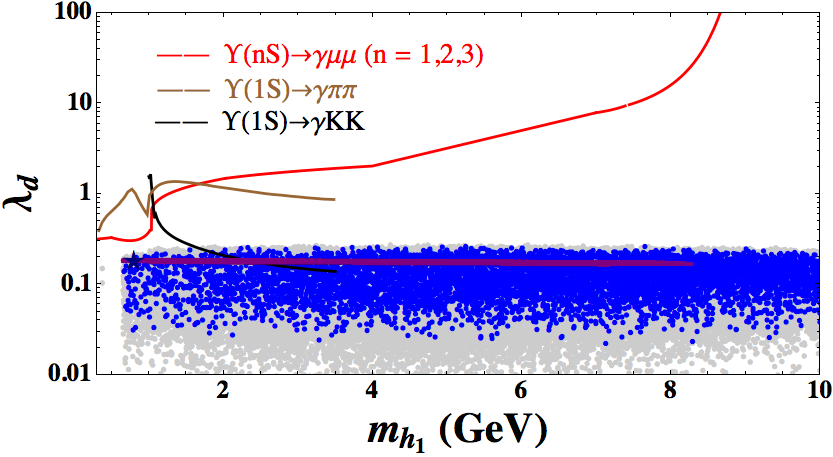}
\caption{Constraints from the decays $h_2\to h_1h_1\to 4f$ (\textit{top}) and from the decays $\Upsilon \to \gamma h_1(h_1\rightarrow\mu\mu,\pi\pi,KK)$ (\textit{bottom}). $\sigma_{4\mu}\equiv\sigma_{h_2}{\rm Br}(h_2\to h_1h_1\to 4\mu)$. To show the constraint from the $2\mu2\tau$ channel on the same plot we convert it into an effective constraint on $4\mu$ by rescaling it with $\frac{{\rm Br}(h_1\to \mu\mu)}{{\rm Br}(h_1\to \tau\tau)}$ (a model-independent quantity). $\lambda_d$ is a tree-level coupling of the down-type interaction $-\frac{\lambda_{d}m_{f_d}}{\sqrt 2 v}h_1\bar{f}_df_d$. Gray and blue points correspond to the gray and blue points in FIG.~\ref{dir_general}. Purple bands correspond to the points in the scan of FIG.~\ref{contour1}.}
\label{upsilonfig}
\end{figure}

Similarly to the light $a_1$ scenario of~\cite{Dermisek:2005ar}, 
relevant constraints may come from the searches for~\cite{Schael:2006cr,Abazov:2009yi} 
\begin{align}
h_2 &\to h_1 h_1, a_1a_1 \to 4b,4\tau,2b2\tau\quad({\rm LEP}),\nonumber\\
h_2 &\to h_1 h_1, a_1a_1 \to 4\mu,2\mu2\tau\quad({\rm Tevatron}).\nonumber
\end{align} 
However, in our case the tree-level couplings of $h_2$ to $h_1h_1$  and $a_1a_1$ are suppressed. This can be seen as follows. Since $h_1$ is strongly singlet-like and $h_2$ is up-type, the coupling $y_{h_2h_1h_1} $ is (for a complete formula, see~\cite{Ellwanger:2009dp})
\begin{align}
y_{h_2h_1h_1} 
&\approx -\frac{\lambda vm_Z\varepsilon }{\sqrt{2} \mu}.
\label{yh2h1h1}
\end{align}
Here we use the mixing parameters at lowest order in $\varepsilon$
\begin{align}
S_{2d} \approx \cot\beta, \ \ 
S_{2s} \approx-\frac{2\varepsilon v m_Z}{m_Z^2+\mu^2}
\end{align}
for moderate $\tan\beta$. Similarly, one can find $y_{h_2a_1a_1}=y_{h_2h_1h_1}$ at this order. Both Br$(h_2\to h_1h_1)$ and Br$(h_2\to a_1a_1)$ are thus suppressed by $\lambda\varepsilon\ll 1$, as is shown in the right column of Fig.~\ref{massfig}. 
(Instead, $h_2$ can dominantly decay into $\chi_1$ and $\chi_2$, while $\chi_2$ dominantly decays into light Higgs bosons and $\chi_1$. These facts imply rich Higgs phenomenology in the DLH scenario and can dramatically change the strategies of searching for the SM-like and light Higgs bosons at colliders~\cite{chi12}.) The asymmetry in Br$(h_2\to h_1h_1)$ w.r.t. $\varepsilon$ is caused by an $\mathcal O(\varepsilon^2)$ correction with the opposite sign of the term in Eq.~(\ref{yh2h1h1}).

\begin{figure}[ht]
\begin{center}
\includegraphics[width=0.36\textwidth]{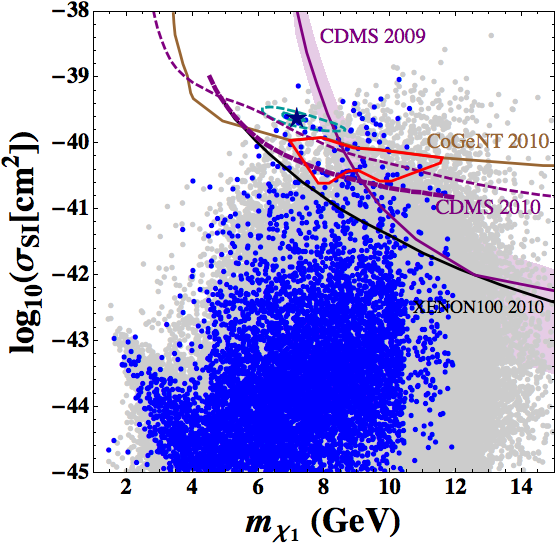}
\caption{Cross section of SI direct detection for $\chi_1$. The scan is over all parameters, in the ranges $0.05\leq\lambda\leq 0.15$, $0.001\leq\kappa\leq0.005$, $|\varepsilon'|\leq 0.25$, $-40\leq A_\kappa\leq 0 \mbox{ GeV}$, $5\leq\tan\beta\leq 50$ and $100\leq\mu\leq250\mbox{ GeV}$. The dark blue (dark) points have a relic density $0.09 \leq \Omega h^2 \leq 0.13$. 
The red contour is the CoGeNT favored region presented in~\cite{Aalseth:2010vx} and the two blue circles are the most recent interpretations of fitting CoGeNT $+$ DAMA/LIBRA~\cite{Hooper:2010uy}. All contours assume a local density which may be sensitive to the relic density. 
The purple, brown, and black lines are the limits from CDMS~\cite{Ahmed:2009zw}, CoGeNT~\cite{Aalseth:2010vx}, and XENON100~\cite{Aprile:2010um}, respectively. Most CoGeNT favored regions have a tension with the CDMS constraints. Consistency between the CoGeNT preferred regions and the XENON100 constraints can be achieved within the scintillation-efficiency uncertainties of liquid xenon~\cite{Hooper:2010uy}.}
\label{dir_general}
\end{center}
\end{figure}

\begin{figure}[ht]
\setlength{\belowcaptionskip}{-20pt}
\begin{center}
\includegraphics[width=0.21\textwidth]{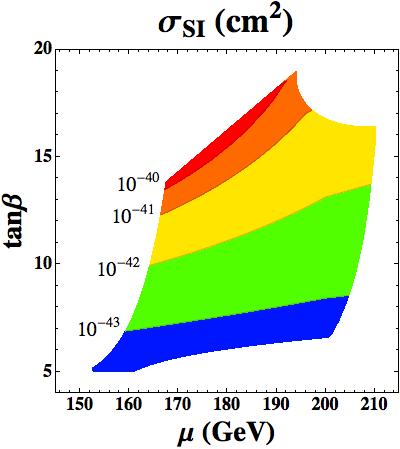}
\includegraphics[width=0.21\textwidth]{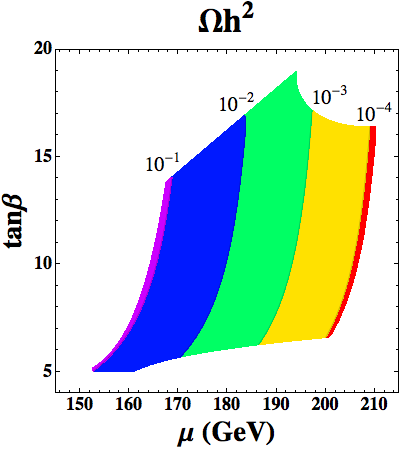}
\includegraphics[width=0.21\textwidth]{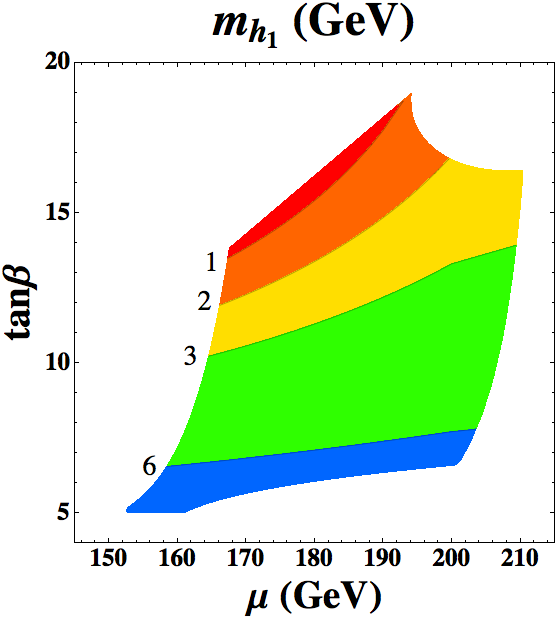}
\includegraphics[width=0.21\textwidth]{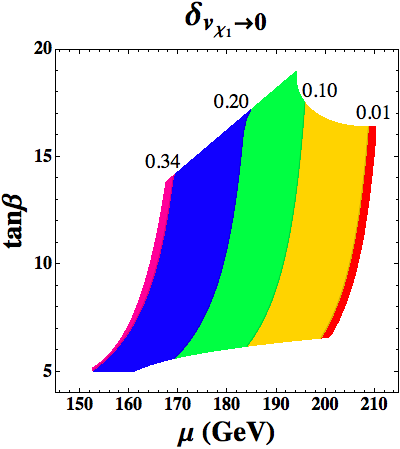}
\caption{Contours of $\sigma_{\rm SI}$ (\textit{top-left}), $\Omega h^2$ (\textit{top-right}), $m_{h_1}$ (\textit{left-bottom}) and $\delta_{v_{\chi_1}\to 0}$ (\textit{right-bottom}) on the $\mu-\tan\beta$ plane, with $\lambda=0.12$, $\kappa=2.7\times 10^{-3}$, $\varepsilon'=0.15$ and $A_\kappa=-24$ GeV. }
\label{contour1}
\end{center}
\end{figure}

The Tevatron constraints from the search for  $h_2\to h_1h_1\to 4f$ are illustrated in the upper panel of FIG.~\ref{upsilonfig}. Almost all points survive. Similar limits from LEP 
are avoided easily for the present parameter values, because $m_{h_2}$ is above the kinematic threshold\footnote{The LEP and Tevatron constraints from the channel $h_2\rightarrow a_1a_1$ are  included in NMSSMTools and in our code, respectively. Points are omitted if the limit is violated. Similarly, the constraint from  $\Upsilon\rightarrow\gamma a_1$ is checked by NMSSMTools, so we present only the limit from $\Upsilon\rightarrow\gamma h_1$ in FIG.~\ref{upsilonfig}. For the numerical results presented in this letter we incorporate {\it{all}} built-in checks in NMSSMTools 2.3.1 (including those from LEP Higgs searches, superpartner searches, $g_{\mu}-2$, flavor physics, $Z$-decay, $\eta_b$ physics, etc.), except the DM relic density. 
The difference between FIG.~\ref{massfig} and FIG.~\ref{dir_general}-\ref{contour1} is that in the latter, $\Omega h^2 \le 0.13$ is also required.}.

$\Upsilon$ physics constrains models with light states through $\Upsilon\rightarrow\gamma (h_1, a_1)\rightarrow \gamma (\mu\mu,\pi\pi,KK)$. Fig.~\ref{upsilonfig} shows the constraints from searches for these decays on the effective coupling $\lambda_d$ of the light state to down-type fermions~\cite{McKeen:2009rm,Aubert:2009cp}. At tree level, 
$\lambda_d\approx \frac{v}{\mu}\big(\lambda +\frac{2\varepsilon\mu}{m_Z}\big)$,
and the scan points typically approach the constrained region only for $\lambda\gtrsim 0.15$. 

$B$-physics may also add non-trivial constraints with a light $a_1$ (e.g., see~\cite{Ellwanger:2009dp}) or $h_1$, because flavor-violating vertices $b(d,s)(a_1,h_1)$ can be generated at loop level. 
These vertices, however, depend strongly on the structure of soft breaking parameters (e.g., see~\cite{Carena:2008ue}). For the input parameters to NMSSMTools used in the scan, the points in the figures are consistent with all $B$-physics constraints including $B_s\rightarrow \mu\mu$, $B_d\rightarrow X_s \mu\mu$, $b\rightarrow s\gamma$, etc. In addition, though not included in NMSSMTools, we also check the constraints from $D$ meson decays (e.g., $D\to l^+ l^-$). Because of the singlet-like nature of $h_1$ and $a_1$, $D$-physics constraints are very weak and can be satisfied easily.

To study the DM physics in the DLH scenario, we perform a second random scan over its parameter region (a narrower region than the one in the first scan). FIG.~\ref{dir_general} shows that the $\chi_1$ DM candidate is characterized by a larger spin-independent direct-detection cross section $\sigma_{\rm SI}$, compared with typical supersymmetric scenarios.  For certain parameter window, 
the correct relic density and a large $\sigma_{\rm SI}$ consistent with the CoGeNT and DAMA/LIBRA preferred region~\cite{Hooper:2010uy} can be simultaneously achieved, and the scenario remains consistent with current experimental bounds (particularly from flavor physics and Higgs searches).  This has been considered difficult or impossible in supersymmetric models~\cite{Feldman:2010ke,Belikov:2010yi,Gunion:2010dy}. 

The large $\sigma_{\rm SI}$ is mainly due to the $h_1-$mediated 
$t-$channel scattering $\chi_1q \to \chi_1 q$, and $\sigma_{\rm SI} \approx$
\begin{align}
\frac{\left (\left(\frac{\varepsilon}{0.04} \right)+ 0.46 \left (\frac{\lambda}{0.1} \right) \left(\frac{v}{\mu}\right) \right)^2 \left(\frac{y_{h_1\chi_1\chi_1}}{0.003}\right)^2 10^{-40} \mathrm{cm}^2}{ \left(\frac{m_{h_1}}{1 \mathrm{GeV}}\right)^4 }. 
\end{align}
The $h_1 \chi_1 \chi_1$ coupling is reduced to $y_{h_1 \chi_1 \chi_1} \approx -\sqrt{2} \kappa$ for a singlino-like $\chi_1$ and singlet-like $h_1$. 
The dependence of $\sigma_{\rm SI}$ on $m_{h_1}^{-4}$ is illustrated in the left panels of FIG.~\ref{contour1}. For the parameter values given in the caption, the LEP search for $h_2\rightarrow bb$ sets the lower boundary of the contoured region, flavor constraints control the upper-right, vacuum stability sets the upper-left limit, and the upper bound on the relic density controls the left and right limits. The sensitivity to $\tan\beta$ enters mainly via $m_{h_1}$.

The $\chi_1$ relic density is largely controlled by the  
$a_1$-mediated annihilation $\chi_1\chi_1 \to f \bar f$, with cross section
\begin{eqnarray}
\sigma_{f \bar f} v_{\chi_1} \approx 
		\frac{3|\, y_{a_1\chi_1\chi_1}\, y_{a_1ff}|^2  (1 -  m_f^2/m_{\chi_1}^2)^{1/2}} {32 \pi  m_{\chi_1}^2 \left (\delta^2+\left |\frac{\Gamma_{a_1} m_{a_1}}{4m_{\chi_1}^2}\right|^2\right)}  ,   
\label{fermanncross}
\end{eqnarray}
where $y_{a_1\chi_1\chi_1} \approx -i \sqrt{2}\kappa $ and $\delta\equiv\left |\frac{1}{1-v_{\chi_1}^2/4}-\frac{m_{a_1}^2}{4m_{\chi_1}^2}\right |$, with $v_{\chi_1}$ denoting the relative velocity of the two $\chi_1$s.). $\delta_{v_{\chi_1}\to0}$ reflects the deviation of $2m_{\chi_1}$ from the $a_1$ resonance.  
In the typical case $m_{a_1}>2m_{\chi_1} > 2 m_b$, the relic density is 
\begin{eqnarray}
\Omega h^2 \approx \frac{0.1\left(\frac{m_{a_1}}{15{\rm GeV}} \right) \left(\frac{\Gamma_{a_1}}{10^{-5} {\rm GeV}} \right) \left(\frac{0.003}{y_{a_1\chi_1\chi_1}}\right)^2 \left(\frac{0.1}{\lambda} \frac{\mu}{v}\right)^2 } {{\rm erfc}\left(\frac{2m_{\chi_1}}{m_{a_1}}\sqrt{x_f\delta_{v_{\chi_1}\to 0}}\right)/{\rm erfc}\left(2.2\right)} \label{rd}
\end{eqnarray}
where $x_f=m_{\chi_1}/T_f$ is the freeze-out point. As a measure of thermal suppression, $\delta_{v_{\chi_1}\to 0}$ enters the complementary error function obtained from the integral over the Boltzmann distribution.  
The inverse dependence of $\Omega h^2$ on $\delta_{v_{\chi_1}\to 0}$ is shown in the right panels of FIG.~\ref{contour1}. Its sensitivity to $\mu$ is mainly through $\delta_{v_{\chi_1}\to 0}$, as $m_{\chi_1}/m_{a_1} \propto \sqrt{\mu}$ for $\tan \beta \simgt 5$. 
To achieve the correct relic density requires $\delta_{v_{\chi_1}\to 0}\approx 0.30-0.35$, which implies $A_\kappa \approx -3.5m_{\chi_1}$, with a tuning range about $\pm0.1m_{\chi_1}$. We emphasize that this process does not generate an antiproton or $\gamma$-ray flux in tension with existing cosmic-ray data because of the Breit-Wigner suppression effect today~\cite{Adriani:2010rc}.

Finally, a benchmark point corresponding to the stars in FIG.~\ref{upsilonfig} and FIG.~\ref{dir_general} is given in Table~\ref{table}. We would like to point out that the chosen set of parameters in the squark, slepton and gaugino sectors in this letter provide a realization of the DLH scenario. Changing them can change the details of the phenomenology, but the basic features will remain intact. We reserve an extended  phenomenological analysis of this scenario for  future work.


\begin{table}[t]
\vglue 0.01cm
\begin{center}
\begin{tabular}{|c|c|c|c|c|c|c|}
\hline
 $\lambda$& $\kappa(10^{-3})$ & $A_\lambda(10^3)$ & $A_\kappa$  & $\mu$  & $\tan\beta$ & $m_{h_1}$ \\
\hline
0.1205 & 2.720 & 2.661  &-24.03&168.0& 13.77 & 0.811\\ \hline
$m_{a_1}$&$m_{\chi_1}$&$m_{h_2}$&${\rm Brhh}$&${\rm Braa}$&$\Omega h^2$& $\sigma_{\rm SI}(10^{-40})$\\ \hline
16.7&7.20&116 &0.158$\%$&0.310$\%$&0.112& 2.34 \\ \hline
\end{tabular}
\end{center}
\caption{Benchmark point. We use the units cm$^2$ for $\sigma_{\rm SI}$ and GeV for dimensionful input parameters, and denote Br$(h_2\to h_1 h_1)$ as Brhh and Br$(h_2\to a_1a_1)$ as Braa. Soft sfermion and gaugino parameters are as given in the caption of FIG.~\ref{massfig}.\label{table}}
\end{table}


\vspace*{-5pt}

\begin{center}
{\bf Acknowledgments}
\end{center}  

\vspace*{-5pt}

Work at ANL is supported in part by the U.S. DOE Grant
DE-AC02-06CH11357.  Work at EFI is supported in part by the DOE Grant DE-FG02- 90ER40560.  
T.L. is supported by the Fermi-McCormick Fellowship and the DOE Grant DE-FG02-91ER40618 at U.~California, Santa Barbara. 
L.-T.W. is supported by the NSF under grant PHY-0756966 and the DOE OJI award under grant DE-FG02-90ER40542.
H.Z. is supported
by the National NSF of China under
Grants 10975004 and the CSC File No. 2009601282. T.L. thanks Princeton U. and Shanghai Jiaotong U. for hospitality during preparation of this work. T.L.  thanks Z.-W. Liu for useful discussions.

\vspace*{-15pt}


\end{document}